\documentclass[useAMS,usenatbib]{mn2e}
\usepackage{txfonts}
\usepackage{longtable}  
\usepackage{rotating}
\usepackage{natbib}
\usepackage{graphicx}
\usepackage{graphics}
\usepackage{psfrag}
\usepackage{amssymb}
\usepackage{multicol}
\usepackage[T1]{fontenc}
\bibpunct{(}{)}{;}{a}{}{,}
\def\Teff{\ensuremath{T_{\mathrm{eff}}}}
\def\logg{\ensuremath{\log g}}

\def\vmic{$\upsilon_{\mathrm{mic}}$}

\def\vsini{\ensuremath{{\upsilon}\sin i}}
\def\kms{$\mathrm{km\,s}^{-1}$}

\def\vr{${\upsilon}_{\mathrm{r}}$}

\def\nlte{non-LTE}
\def\llm{{\small LLMODELS}}

\def\synth{{\small{SYNTH}}3}
\def\ubvri{\ensuremath{UB\,VRI_{c}}}
\def\ubv{\ensuremath{UB\,V}}
\def\bz{\ensuremath{\langle B_z\rangle}}

\title[A probable PMS CP star in Stock~16]{A probable pre-main sequence chemically peculiar star in the open cluster Stock~16}

\author[M. Netopil et al.]{M. Netopil$^{1}$\thanks{E-mail:
mnetopil@physics.muni.cz}, L. Fossati$^{2}$, E. Paunzen$^{1}$, K. Zwintz$^{3}$, O. I. Pintado$^{4}$ and S. Bagnulo$^{5}$\\
$^{1}$Department of Theoretical Physics and Astrophysics, Masaryk University, Kotl\'a\v{r}sk\'a 2, 611 37 Brno, Czech Republic\\
$^{2}$Argelander-Institut f\"{u}r Astronomie der Universit\"{a}t Bonn, Auf dem H\"{u}gel 71, 53121 Bonn, Germany\\
$^{3}$Instituut voor Sterrenkunde, KU Leuven, Celestijnenlaan 200D, 3001 Leuven, Belgium\\
$^{4}$Instituto Superior de Correlaci\'on Geol\'ogica, CONICET, Miguel Lillo 205, 4000 San Miguel de Tucum\'an, Argentina\\
$^{5}$Armagh Observatory, College Hill, Armagh BT61 9DG, Northern Ireland, UK}

\begin{document}

\date{}

\pagerange{\pageref{firstpage}--\pageref{lastpage}} \pubyear{2014}

\maketitle

\label{firstpage}

\begin{abstract}
We used the Ultraviolet and Visual Echelle Spectrograph of the ESO-Very Large Telescope to obtain a high resolution and high signal-to-noise ratio spectrum of Stock~16-12, an early-type star which previous $\Delta$a photometric observations suggest being a chemically peculiar (CP) star. We used spectral synthesis to perform a detailed abundance analysis obtaining an effective temperature of 8400$\pm$400\,K, a surface gravity of 4.1$\pm$0.4, a microturbulence velocity of 3.4$^{+0.7}_{-0.3}$\,\kms, and a projected rotational velocity of 68$\pm$4\,\kms. We provide photometric and spectroscopic evidence showing the star is most likely a member of the young Stock 16 open cluster (age 3--8\,Myr). The probable cluster membership, the star's position in the Hertzsprung-Russell diagram, and the found infrared excess strongly suggest the star is still in the pre-main-sequence (PMS) phase. We used PMS evolutionary tracks to determine the stellar mass, which ranges between 1.95 and 2.3\,M$_\odot$, depending upon the adopted spectroscopic or photometric data results. Similarly, we obtained a stellar age ranging between 4 and 6\,Myr, in agreement with that of the cluster. Because the star's chemical abundance pattern resembles well that known of main sequence CP metallic line (Am) stars, the object sets important constraints to the diffusion theory. Additional spectroscopic and spectropolarimetric data allowed us to conclude that the object is probably a single non-magnetic star. 

\end{abstract}

\section{Introduction}

The classical metallic-line stars of the upper main sequence (often denoted as Am/Fm or CP1 stars), are A- to early F-type objects \citep{Pres74}. These stars, as such, were first described by \citet{Titu40} using spectra of Hyades cluster members. In general, Am stars are characterized by overabundances of metallic elements heavier than iron, whereas calcium, scandium, carbon, nitrogen and oxygen are underabundant.

There are two different models based upon atomic diffusion to explain the Am phenomenon. \citet{Wats71} proposed that the separation of chemical elements occurred just below the hydrogen convection zone where calcium has a small radiative acceleration. This implies that on the main sequence (MS) a very small mass fraction has anomalous abundances. In the subgiant phase, the anomalies will be diluted due to the increased mixing. In the other model, proposed 
by \citet{Rich00}, the separation occurs much deeper in the stellar interior. Therefore, on the MS, a much larger mass fraction has anomalous abundances. As a consequence, the chemical peculiarities remain for a longer time as the star evolves on the subgiant branch: the abundance anomalies remain visible till convection dominates.

The chemical separation resulting from atomic diffusion can affect both the surface and interior of pre-main sequence (PMS) stars \citep{Vick11} and the mass in the different convection zones and internal elemental concentrations can be already modified before a star arrives on the MS. The amplitude of the internal abundances depends on the stellar mass; equivalently, the age at which abundance anomalies appear at the stellar surface also depends on the stellar mass. In the presence of weak mass loss, and no turbulent mixing, rotation or magnetic fields, significant internal variations and surface
anomalies can appear already after about 2\,Myr in early A-type stars, and at about 20\,Myr in cooler F-type stars. To constrain these models it is essential to know when the Am phenomenon first appears on the PMS, but a PMS Am star has not yet been found so far.

For another group among the chemically peculiar (CP) stars (the magnetic Bp/Ap objects), there were already efforts to identify representatives on the PMS. However, up to now only one possible candidate with weak Ap/Bp peculiarities (V380 Ori A) was detected \citep{folsom2012}. Recently, \citet{Bailey14} presented the time-dependent evolution of chemical abundances for (MS) cluster Bp stars. A similar approach for Am objects and the detection of a PMS prototype is important to provide further observational constraints to the diffusion theory and on the time-scale of the evolution of the chemical abundances of CP stars. In order to search for PMS CP stars, observations of open clusters are preferable because they contain samples of stars of well constrained age and homogeneous initial chemical composition, characteristics suited for the study of processes linked to stellar structure and evolution.

The paper is arranged as follows. In Section \ref{targetstar} we describe the open cluster and the target star. A detailed spectroscopic analysis of the target star is presented in Section \ref{sect:specanalysis}. These results are essential to provide further constraints on the cluster membership and evolutionary stage (Section \ref{sect:clustermember}). In Section \ref{sect:discussion} we discuss the results, and Section \ref{sect:conclusion} concludes the paper.

\section{Target star and host open cluster}
\label{targetstar}
\begin{figure}
\begin{center}
\includegraphics[width=85mm,clip]{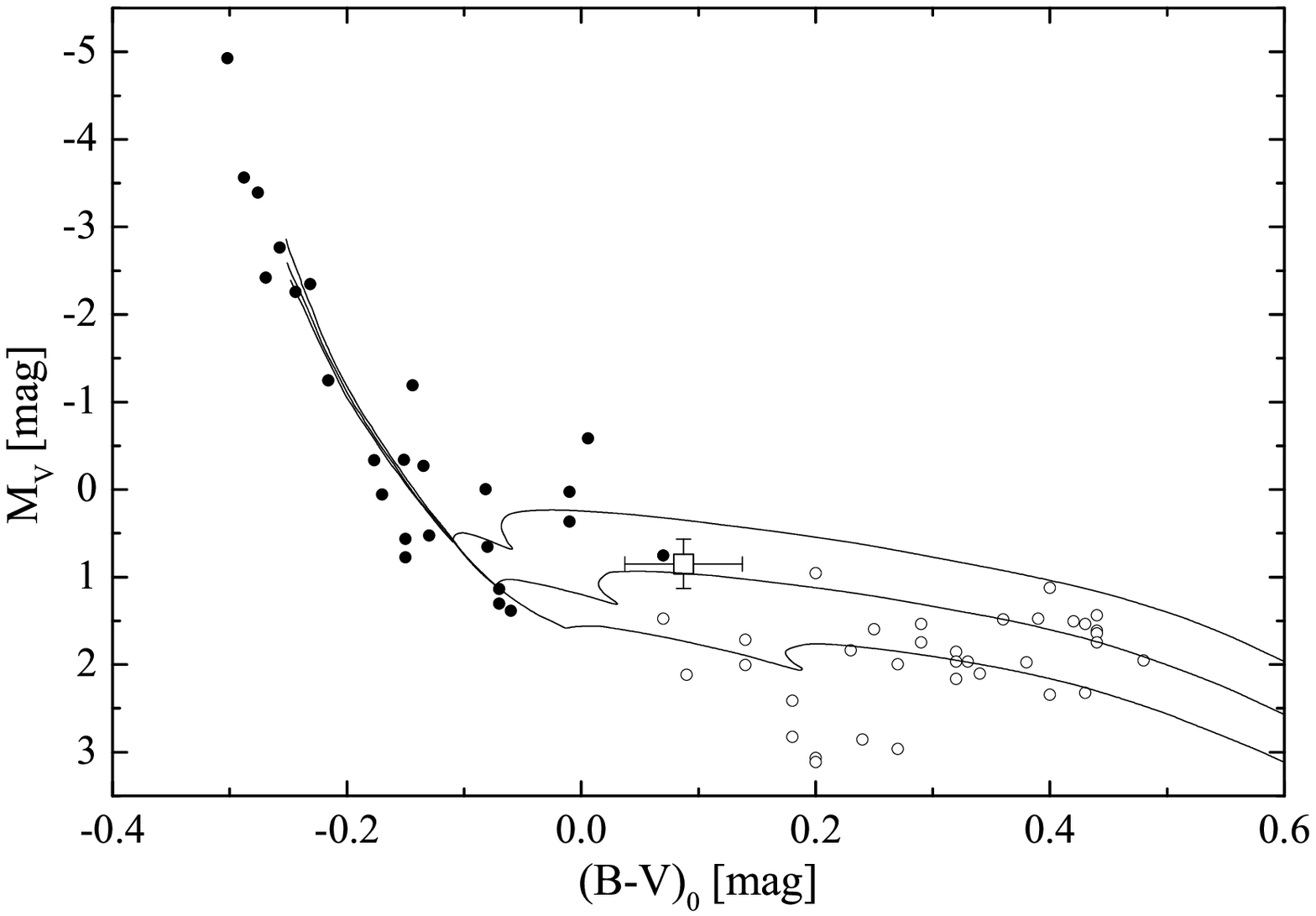}
\caption{The CMD of Stock~16 based on member stars listed by \citet{turner1985} and \citet{vazquez2005}. Probable PMS objects identified by \citet{vazquez2005} are indicated by open circles, while the target star Stock~16-12 is marked by an open square. Whenever applicable, the stars were individually dereddened as described in Section \ref{sect:clustermember}, otherwise the average $E(B-V)=0.51$~mag was applied. The average cluster distance of 1910\,pc was used to derive the absolute magnitudes. The isochrones for 3, 5, and 8\,Myr by \citet{Bressan12}, which include the PMS phase, are shown as solid lines. }
\label{fig:cmd} 
\end{center} 
\end{figure}

The young open cluster Stock\,16 is located in the H\,{\sc ii} region RCW~75, which furthermore appears to be part of the extended associations Cen~OB1 and Cen~R1 \citep{turner1985}. The first detailed investigation of the cluster was carried out by \citet{fenkart1977} in the photographic \textit{RGU} system, followed by the photoelectric $\ubv$ study by \citet{turner1985}. Later on, \citet{vazquez2005} presented a deep CCD multi-colour investigation and identified several PMS objects. Finally, \citet{paunzen2005} studied the cluster in the $\Delta a$ photometric system as part of a survey to detect CP stars. All mentioned studies agree in the obtained cluster parameters. Using the individual results, we derived the average values $E(B-V)=0.51\pm0.02$~mag for the reddening and 1910$\pm$85\,pc for the distance. Since the cluster's metallicity is not known, which strongly influences the distance determination, the uncertainty on the distance is very probably underestimated. We therefore adopt a more realistic value of about 10 per cent ($\sim$\,200\,pc). In the literature \citep{turner1985, paunzen2005, vazquez2005} the cluster age is estimated to be between 3 and 8\,Myr. Hence, at such a young age, cluster stars with spectral type A and later are most likely still in their PMS stage. Fig.~\ref{fig:cmd} shows the dereddened colour--magnitude diagram (CMD) of the cluster.  

Note that the cluster position given in SIMBAD and in the open cluster data base WEBDA\footnote{\tt http://webda.physics.muni.cz/} \citep{mp2003} disagrees by about 10 arcmin with the one used by all the aforementioned studies. Because of this offset, \citet{santos2012} obtained for the cluster a rather low stellar density contrast to the background. However, this study derived values for the age and the distance similar to the ones of Stock~16 listed above, but a reddening lower by $\sim$\,0.2\,mag. Nevertheless, this provides a further proof of a young area, which probably exhibits strong differential reddening.

The OB association Cen OB1 was studied by \citet{kaltcheva94} based on Str\"omgren photometry for two dozen stars. They obtained a distance of $\sim$\,2.3\,kpc, an age of about 9\,Myr, and a strong differential reddening in $E(B-V)$ between 0.3 and 1.1\,mag, not uncommon for a dusty region. \citet{corti2013} derived a somewhat larger distance of $\sim$\,2.6$\pm$0.5\,kpc, an average radial velocity of \vr=$-$20.0\,\kms\ and an average proper motion of $-$4.78$\pm$0.10\,/\,$-$0.93$\pm$0.10 mas yr$^{-1}$. The latter is identical to the one listed by \citet{zejda2012} for Stock~16 ($-$5.0$\pm$0.4 / $-$0.9$\pm$0.4 mas yr$^{-1}$). A much closer distance of 1.8$\pm$0.4\,kpc was instead found by \citet{kaltcheva2013}. They attribute the large difference to an overestimation of the absolute magnitude when using spectral types and standard relations. 

Hence, as already pointed out by \citet{turner1985}, there is a strong evidence that Stock~16 is a member of an extended structure, which includes at least one further confirmed young open cluster \citep[NGC 4755;][]{corti2013} with parameters nearly identical to that of Stock~16.

The photometric study by \citet{paunzen2005} detected a probable CP star in the area of Stock~16 (Stock~16-12: V$\sim$13.4\,mag, J2000 RA 13:18:51, Dec. $-$62:29:22) due to a positive peculiarity index ($\Delta a$\,=\,+24\,mmag). Such a value is in general indicative of characteristics typical of magnetic CP stars \citep{paunzen2005b}. Later, the star was classified as A2\,Si with strong metals by \citet{paunzen2011} based on a low-resolution spectrum (2\,\AA\,pixel$^{-1}$). The problems arising in the classification of the various peculiar groups are extensively discussed by \citet{jaschek1974}, and one also can not exclude a bias in the classification process due to the measured $\Delta a$. However, the noticeable strong metals are already a hint for a later type star and hence a possible nature as metallic-line (Am) star as found in Section \ref{sect:specanalysis}. Only some representatives of this group show positive $\Delta a$ values, but \citet{paunzen2005} used the apparent and not intrinsic colours for the diagnostic $\Delta a$ diagram. The somewhat lower reddening of Stock~16-12 compared to the cluster's mean (see discussion in Section \ref{sect:clustermember}) shifts the star to redder colours in the diagram, resulting in an insignificant $\Delta a$ value.

We obtained an additional spectrum of this star with the 2.15m telescope at the Complejo Astron{\'o}mico el Leoncito (CASLEO) on 2012 May 11 using the same configuration as given by \citet{paunzen2011}, except for a small shift in the blaze angle. With the new information provided in the present paper, we would classify Stock~16-12 as kA2\,hA3\,mA7 based on the Ca {\sc ii} K, hydrogen, and metallic lines. There are no noticeable differences between the two available low resolution spectra, though both exhibit a low signal-to-noise ratio (S/N) of $\sim$\,25. We therefore decided to acquire an additional higher quality spectrum for a detailed analysis, which is presented in Section \ref{sect:specanalysis}.

\section{Spectroscopic analysis}
\label{sect:specanalysis}
We obtained one spectrum of Stock\,16-12 on 2013 May 20 with the Ultraviolet and Visual Echelle Spectrograph (UVES) of the ESO Very Large Telescope (VLT). This is a cross-dispersed \'echelle spectrograph, which in the standard mode and the adopted configuration (1 arcsec slit width, red arm, and grating \#520) yields a resolving power of 40\,000. With exposure time of 3000\,s, at an airmass of 1.35 and a seeing of about 0.9 arcsec, we obtained a S/N pixel$^{-1}$ at $\lambda\sim$5500\,\AA\ of 210. 

All reduction steps were performed within the UVES pipeline (version 5.2.0) and {\small{REFLEX}}\footnote{\tt http://www.eso.org/sci/software/reflex/}. Neither binning
nor smoothing was applied to the \'echelle spectrum.
The spectrum, normalized by fitting a low order polynome to carefully 
selected continuum points, covers the wavelength range 4170--6200\,\AA, with 
a 60\,\AA\ wide gap at about 5200\,\AA.

To analyse the spectrum, we computed model atmospheres of Stock\,16-12 using the \llm\ stellar model atmosphere code \citep{llm}. For all the calculations 
local thermodynamical equilibrium (LTE) and plane-parallel geometry were 
assumed. Convection was implemented according to the \citet{cm1,cm2} model of convection \citep[see][for more details]{heiter}. We used the Vienna Atomic Line Database \citep[VALD;][]{vald1,vald2,T83av} as a source of atomic line parameters for opacity calculations and abundance analysis.

We measured the radial velocity (\vr) and the projected rotational velocity (\vsini) by fitting synthetic spectra, calculated with \synth\ \citep{synth3}, to the observed profiles of weakly blended lines, obtaining \vr=$-$18.0$\pm$1.0\,\kms\ and \vsini=68.0$\pm$4.0\,\kms.

\begin{figure}
\begin{center}
\includegraphics[width=85mm,clip]{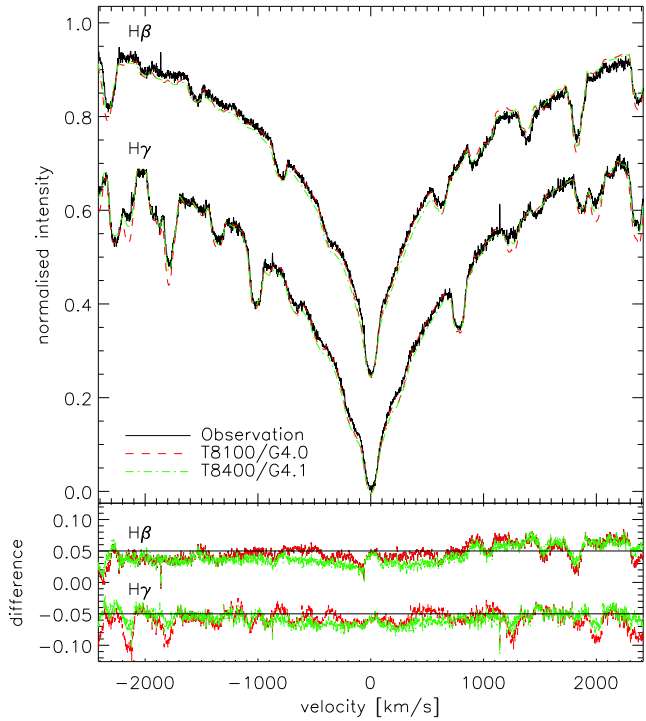}
\caption{
Top panel: comparison between the observed H$\gamma$ and H$\beta$ line profile (black solid line) and synthetic profiles calculated with the final adopted parameters (dash-dotted green line) and with the parameters which were found to best fit the hydrogen line profiles (dashed red line). The H$\gamma$ line profile was rigidly shifted downwards by 0.2. The wavelength dimension has been converted in velocity and centred at the laboratory wavelength of the H$\beta$ and H$\gamma$ lines, respectively. Bottom panel: difference between the synthetic and observed H$\gamma$ and H$\beta$ line profiles rigidly shifted respectively downwards and upwards by 0.05. The line colours and styles are as for the top panel.}
\label{fig:hydrogen} 
\end{center} 
\end{figure}

To determine the fundamental atmospheric parameters we used both hydrogen and metallic lines. To spectroscopically estimate the effective temperature (\Teff) and surface gravity (\logg) from hydrogen lines, we fitted synthetic line profiles to the observed H$\gamma$ and H$\beta$ lines. The adopted spectral synthesis code, \synth, incorporates the algorithm by \citet{barklem2000}\footnote{\tt http://www.astro.uu.se/$\sim$barklem/hlinop.html} which takes into account not only self-broadening but also Stark broadening (see their section 3). The fit to the two covered hydrogen lines led to \Teff=8100$\pm$400\,K and \logg=4.0$\pm$0.4~dex (see Fig.~\ref{fig:hydrogen}). The rather large error bars are due to uncertainties in the normalization and to the small, thought not negligible, sensitivity of the hydrogen line wings to gravity variations.

The metallic-line spectrum provides further constraints on the atmospheric 
parameters. The presence of rather strong line blending, due to \vsini\ and to the overabundance of various elements, did not allow us to measure spectral lines with equivalent widths. We therefore analysed the metallic lines using synthetic spectra. We first selected the least blended metallic lines with a well defined continuum level and then, for a given set of atmospheric parameters (\Teff, \logg, and microturbulence velocity -- \vmic), fit the abundance of each selected line using the tools described in \citep{fossati2007,fossati2008}. We repeated this operation for a set of atmospheric parameters ranging from 8000 to 9000\,K in \Teff, from 3.6 to 4.5~dex in \logg, and from 2.0 to 5.0\,\kms\ in \vmic. We adopted steps of 100\,K in \Teff, 0.1\,dex in \logg, and 0.1\,\kms\ in \vmic. For each given set of atmospheric parameters and abundances we then calculated a synthetic spectrum covering the observed wavelength range. We finally determined the overall (excluding the regions covered by the hydrogen lines and the Na\,{\sc i}\,D lines at $\lambda\sim$5892\,\AA\ affected by strong \nlte\ effects) best fitting synthetic spectrum, hence set of atmospheric parameters and abundances, using $\chi^2$ statistics. 

In the considered temperature regime and in particular for the spectral region bluewards of H$\beta$, the continuum level is extremely sensitive to temperature variations. This would lead to the introduction of a bias in the $\chi^2$ minimization if applied using an observed spectrum with a fixed normalization, i.e. the $\chi^2$ would be driven mostly by the fit to the continuum level, rather than by the fit to the metallic lines. To avoid this problem we renormalized the observed spectrum in order to perfectly match the continuum level of each calculated synthetic spectrum. In this way, variations of the $\chi^2$ are due exclusively to the quality of the fit of the single spectral features. Using this technique we obtained \Teff=8700$\pm$200\,K, \logg=4.2$\pm$0.4\,dex, and \vmic=3.7$\pm$0.3\,\kms.

The two adopted parameter indicators did not lead to consistent effective temperature values. This could be due to various systematics affecting the determination of the atmospheric parameters (\Teff\ in particular) with the adopted indicators. In particular hydrogen lines recorded with \'echelle spectra are usually difficult to normalize. In this case, the normalization of the hydrogen lines was particularly uncertain due to the `wavy' shape of the not-normalized reduced spectrum. In addition, \citet{fossati2011} showed that a determination of the atmospheric parameters based on the abundance fit of metallic lines might lead to an overestimation of the effective temperature of up to a few hundreds of Kelvin. On the basis of these considerations we finally set the effective temperature at \Teff=8400$\pm$400\,K and the surface gravity at \logg=4.1$\pm$0.4\,dex. Having fixed \Teff\ and \logg\ we then used the metallic lines and the method described above to determine the best-fitting \vmic, obtaining \vmic=3.4$^{+0.7}_{-0.3}$\,\kms.

The only way to solve the problem of the discordant effective temperature values would be to use further parameter indicators, such as photometry, which could be directly converted to fundamental parameters (\Teff\ and \logg) or calibrated to physical units to fit the spectral energy distribution (SED). Unfortunately, both methods are strongly affected by interstellar reddening, which for the Stock\,16 open cluster is not known well enough to ensure an effective temperature determination more reliable than that provided by the hydrogen and metallic lines. Furthermore, a slight differential reddening in the young cluster results in a large uncertainty if using the mean reddening of the cluster.

\begin{figure}
\begin{center}
\includegraphics[width=85mm,clip]{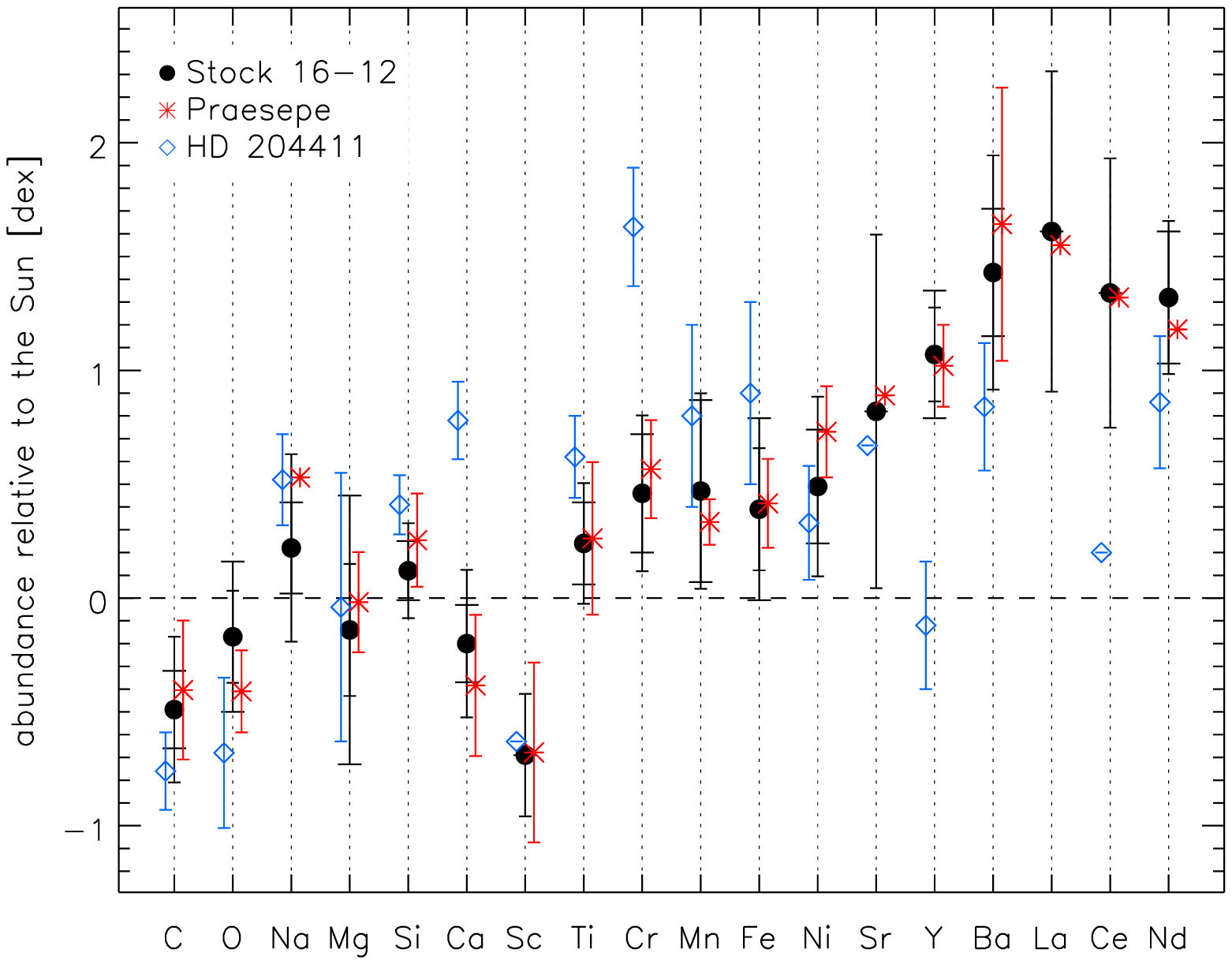}
\caption{Element abundance relative to the Sun of the Stock\,16-12 atmosphere (full black points) in comparison to the average abundance (red asterisks) of the Am stars member of the Praesepe open cluster \citep{fossati2007,fossati2008} and of the Ap star HD\,204411 which has an effective temperature similar to that of Stock\,16-12 \citep{ryabchikova2005}. For all objects the same solar abundance set \citep{asplund2009} was used. Each abundance value of Stock\,16-12 is shown with two uncertainty values: the standard deviation from the mean (column 2 of Table~\ref{tab:abundances}; assumed to be 0.2\,dex for the elements analysed from just one line) and the uncertainty obtained by adding in quadrature the standard deviation from the mean and the variation in abundance due to an increase of 400\,K in effective temperature (column 3 of Table~\ref{tab:abundances}).} 
\label{fig:abundances} 
\end{center} 
\end{figure}

We finally measured the abundance of 18 elements (see Table~\ref{tab:abundances}) obtaining an abundance pattern which well resembles that of CP metallic-line stars (Am stars). Fig.~\ref{fig:abundances} shows the derived abundance pattern, relative to that of the Sun \citep{asplund2009}, in comparison with the average abundance of the Am stars member of the Praesepe open cluster \citep{fossati2007,fossati2008} and of the Ap star HD\,204411 which has an effective temperature similar to that of Stock\,16-12 \citep{ryabchikova2005}. The abundance pattern of Stock\,16-12 presents all classical chemical signatures of Am stars: underabundances of C, N, O, Ca\footnote{Note that an underabundance of Ca is not necessarily present in Am CP stars.}, and Sc, as well as overabundances of Fe-peak and rare-earth elements. Stock\,16-12 shows also a slight overabundance of Na, which is not classically considered as a CP element in Am stars, though \citet{fossati2007} found it to be systematically overabundant in each analysed Am star, and \citet{takeda2012} found it to directly correlate with Fe in early F- and A-type stars. Note that we determined the sodium abundance using the lines at $\lambda\sim$5688\,\AA\ which are very little affected by \nlte\ effects; we did not use the Na\,{\sc i}\,D lines at $\lambda\sim$5892\,\AA\ which would instead lead to a strong Na overabundance due to \nlte\ effects. The strong blending did not allow us to measure the abundance of other elements which are known to be overabundant in Am stars (i.e., V, Co, Cu, Zn, Zr, Pr, Sm, Gd, Dy, Ho, Er, and Yb). Nevertheless, the observed spectrum would still be well fitted when assuming an overabundance of 0.5\,dex for V, Co, Cu, Zn, and Zr and of 1.5\,dex for Pr, Sm, Gd, Dy, Ho, Er, and Yb, as expected for Am stars.

\begin{table}
\caption[ ]{LTE atmospheric abundances of Stock\,16-12 with the error estimates based on the internal scattering from the number of analysed lines,
$n$. The third column gives the variation in abundance estimated by increasing \Teff\ by 400\,K. The fifth column gives the abundances of Stock\,16-12 relative to the solar values from \citet{asplund2009}. The last column lists the abundances of the solar atmosphere calculated by \citet{asplund2009}.}
\label{tab:abundances}
\begin{center}
\begin{tabular}{l|ccc|c|c}
\hline
\hline
Ion & \multicolumn{4}{|c|}{Stock\,16-12} &  Sun \\                    
    & $\log (N/N_{\rm tot})$ & $\Delta$\Teff &  $n$ & [$N_{el}/N_H$] & $\log (N/N_{\rm tot})$ \\     
\hline                                                                                    
C  &  $-$4.10          & 0.25 &  1 & $-$0.49 &  $-$3.61 \\ 
O  &  $-$3.52          & 0.03 &  1 & $-$0.17 &  $-$3.35 \\ 
Na &  $-$5.58          & 0.36 &  1 &	+0.22 &  $-$5.80 \\ 
Mg &  $-$4.58          & 0.21 &  1 & $-$0.14 &  $-$4.44 \\ 
Si &  $-$4.41          & 0.06 &  1 &	+0.12 &  $-$4.53 \\ 
Ca &  $-$5.90$\pm$0.05 & 0.32 &  3 & $-$0.20 &  $-$5.70 \\ 
Sc &  $-$9.58          & 0.18 &  1 & $-$0.69 &  $-$8.89 \\ 
Ti &  $-$6.85$\pm$0.26 & 0.05 &  2 &	+0.24 &  $-$7.09 \\ 
Cr &  $-$5.94$\pm$0.04 & 0.34 &  2 &	+0.46 &  $-$6.40 \\ 
Mn &  $-$6.14          & 0.38 &  1 &	+0.47 &  $-$6.61 \\ 
Fe &  $-$4.15$\pm$0.12 & 0.24 & 41 &	+0.39 &  $-$4.54 \\ 
Ni &  $-$5.33          & 0.34 &  1 &	+0.49 &  $-$5.82 \\ 
Sr &  $-$8.35          & 0.75 &  1 &	+0.27 &  $-$9.17 \\ 
Y  &  $-$8.76$\pm$0.08 & 0.19 &  2 &	+1.07 &  $-$9.83 \\ 
Ba &  $-$8.43$\pm$0.23 & 0.46 &  2 &	+1.43 &  $-$9.86 \\ 
La &  $-$9.33$\pm$0.18 & 0.68 &  2 &	+1.61 & $-$10.94 \\ 
Ce &  $-$9.12$\pm$0.04 & 0.59 &  3 &	+1.34 & $-$10.46 \\ 
Nd &  $-$9.30$\pm$0.13 & 0.31 &  2 &	+1.32 & $-$10.62 \\ 
\hline											  
\Teff     & \multicolumn{4}{|c|}{8400\,K} & 5777\,K \\			
\logg     & \multicolumn{4}{|c|}{4.1\,dex}     & 4.44\,dex    \\				 
\hline											  
\end{tabular}
\end{center}
\end{table}

For stars in the temperature regime of Stock\,16-12, variations on the effective temperature have the largest impact on abundance values \citep[see e.g.,][]{fossati2009}. For this reason, and given the large difference in \Teff\ obtained using hydrogen and metal lines, it is important to check the abundance pattern still resembles that of an Am star when accounting for the uncertainty on \Teff. The abundances of Stock\,16-12 shown in Fig.~\ref{fig:abundances} present two error bars, where the larger one takes into account also the uncertainty due to an increase in \Teff\ by 400\,K. Even with the larger uncertainty the abundance pattern of Stock\,16-12 resembles well that of Am stars, confirming therefore its classification.

The rather large \vmic\ value we derived from the UVES spectrum is in line with that typical of Am stars \citep{landstreet1998,landstreet2009}. Nevertheless, the behaviour of the microturbulence velocity in very young (PMS) stars is not yet understood. Abundance analysis of young stars in the same temperature regime of Stock\,16-12 shows an average \vmic\ value between 2.5 and 3\,\kms\ \citep{folsom2012,zwintz2013}, slightly enhanced compared to that of MS stars of the same temperature, though somewhat smaller than what we obtained for Stock\,16-12. This further strengthens the conclusion that Stock\,16-12 is a CP Am star.

However, CP Am stars are commonly found in close binary systems \citep{abt1985,debernardi2000}. To check whether Stock\,16-12 is member of a close binary, we derived the star's radial velocity from an interrupted UVES exposure, obtained within the same service mode observing programme on 2013 April 28 (exposure time 722\,s; S/N pixel$^{-1}$ at $\lambda\sim$5500\,\AA\ of 49), and two further UVES spectra obtained within a Director's Discretionary Time (DDT) programme on 2014 April 12 with exposure times of 1205\,s and a S/N of 56 and 52, respectively. Using the same method as applied on the high S/N spectrum, we measured a consistent radial velocity of \vr=$-$18.0$\pm$1.0\,\kms. Thus, we conclude that Stock\,16-12 is probably not in a binary system, unless it is a wide binary with a period of several years.

To gather more information on the CP nature of the star, we obtained Focal Reducer and Low Dispersion Spectrograph 2 (FORS2) spectropolarimetric observations to attempt the detection of a large scale magnetic field. The FORS2 low-resolution spectropolarimeter \citep{app1992,app1998} is attached to the Cassegrain focus of the 8-m Antu telescope of the ESO VLT of the Paranal Observatory. The observations were performed using the 2k $\times$ 4k MIT CCDs (pixel size 15\,$\mu$m\,$\times$\,15\,$\mu$m) and a slit width of 1.0 arcsec in order to collect more photons and increase the S/N. We also adopted the 200\,kHz/low/1 $\times$ 1 readout mode and the GRISM\,1200B. Each spectrum covers the 3700--5100\,\AA\ spectral range which includes most of the Balmer lines, except H$\alpha$, and a number of metallic lines. The star was observed once on 2014 April 18 with a sequence of spectra obtained rotating the quarter waveplate as follows: $-$45$^{\circ}$, $+$45$^{\circ}$, $+$45$^{\circ}$, $-$45$^{\circ}$. Each of the four spectra was obtained with an exposure time of 1568 s which led to a Stokes \textit{I} spectrum with a S/N pixel$^{-1}$ calculated at 4950\,\AA\ of about 1120. We reduced and analysed the FORS2 data using the routines and technique described in \citet{bagnulo2012} and references therein, without detecting a magnetic field: by using the whole spectrum we obtained an average longitudinal magnetic field value \bz\ of $-$11$\pm$76\,G, while using the hydrogen lines we obtained \bz\ = $-$193$\pm$129\,G. Hence, there is no evidence for a magnetic field with a strength typical of that of 
Ap stars.

\section{Cluster membership and evolutionary status}
\label{sect:clustermember}

Three previous studies list the object analysed here as member of Stock~16 \citep[][]{fenkart1977,turner1985,paunzen2005}, whereas \citet{vazquez2005} denoted the star as probable non-member by means of CCD $\ubvri$ photometry. This last classification is probably due to the deviation from the cluster sequence in the $(U-B)/V$ CMD, since in all other colours no disagreement is noticeable. However, compared to the photoelectric $(U-B)$ measurement by \citet{turner1985}, the CCD study provided a much redder colour by 0.14\,mag. This can be attributed to well known difficulties in reproducing the standard $(U-B)$ colour with CCDs \citep[see e.g.,][]{sung2000}. We therefore adopt the photoelectric measurement for this colour. The remaining colours in common between the investigations are in excellent agreement, also if compared with measurements by the AAVSO Photometric All-Sky Survey [APASS,\footnote{http://www.aavso.org/apass} Data Release 7 (DR7)], which provides data in the $B\,Vg'r'i'$ filters.

Since the cluster is located at a distance of $\sim$\,1.9\,kpc, proper motion as membership criterion is in general not that efficient anymore compared to nearby open clusters. Nevertheless, the average cluster motion given by \citet{zejda2012} is in agreement with that listed in the UCAC4 \citep{zacharias2013} and PPMXL \citep{Roes10} catalogues: $-$5.7/3.0 ($\pm$2.7) and $-$5.7/5.5 ($\pm$9.8) mas yr$^{-1}$, respectively. The errors for both directions are given in parentheses. By employing the method described by \citet{Bala98}, which takes into account the individual errors, we derived kinematic membership probabilities for Stock~16-12 of 62 and 91 per cent using UCAC4 and PPMXL data, respectively.    

The interstellar reddening is a further useful information to constrain membership. With the adopted spectroscopic stellar parameters and their uncertainties from Section \ref{sect:specanalysis}, the empirical colour-temperature relation by \citet{worthey2011}, and the observed colours, we determined $E(B-V)=0.36\pm0.05$~mag. The various colour excess ratios are well in line with a standard reddening law $R_V=3.1$, which was also found by \citet{vazquez2005} for the whole cluster. As comparison, we applied the $Q$-method \citep[e.g.,][]{Guti75} to the photoelectric $\ubv$ data by \citet{turner1985} using stars in Stock~16 brighter than $V$=13\,mag, which are most likely earlier than spectral type A0 and reached already the MS. On the basis of 14 stars, we obtained a reddening range of $0.40 < E(B-V) < 0.56$~mag. Hence, within the errors, the object lies on or slightly below the lower reddening border. On the other hand, the analysis by \citet{kaltcheva2013} of the surrounding Cen OB1 association shows that the bulk of foreground objects exhibits a reddening lower than the one of Stock~16-12. A detailed investigation of the cluster area with Str\"omgren photometry would be helpful to further constrain both reddening and membership.

Unfortunately, the derived radial velocity for the target star (\vr=$-$18.0$\pm$1.0~\kms, see Section \ref{sect:specanalysis}) does not provide an additional membership indication. \citet{kharch2005} list an average velocity of $-42 \pm 9$\,\kms using literature data for five O- and early B-type stars in the range of $-$21 to $-$74~\kms. However, according to \citet{sana2012} more than 70 per cent of massive stars are part of binary systems. Indeed, most of these five objects are mentioned in the literature as radial velocity variables \citep[e.g. by][]{crampton1972}. Nevertheless, the derived radial velocity is in very good agreement with the average value of the Cen OB1 association \citep[\vr=$-$20.0\,\kms;][]{corti2013}. Furthermore, the Na\,D lines in the spectrum show massive interstellar absorption lines, which would agree with the membership of the star in a young cluster or association, which still holds part of the original molecular cloud. In a few years the precise \textit{Gaia} data will allow one to obtain an unambiguous membership determination.

\begin{figure}
\begin{center}
\includegraphics[width=85mm,clip]{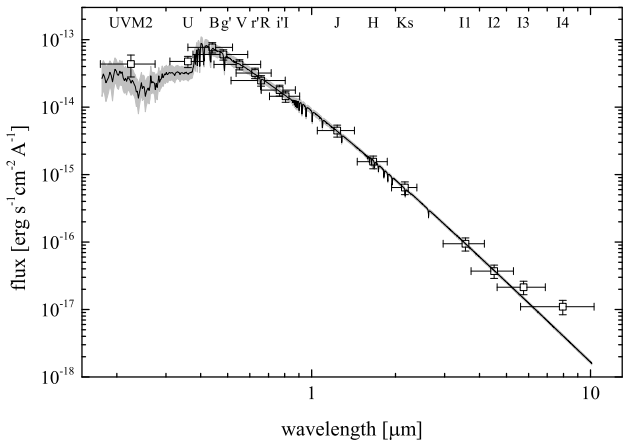}
\caption{The SED of Stock~16-12 using the available photometric measurements in the UVOT, Johnson, SDSS, 2MASS, and IRAC bands (squares) compared to theoretical fluxes of the adopted model atmosphere (8400\,K, black line). The grey area represents the differences to models with temperatures $\pm$400\,K to indicate the changes in the flux. Furthermore, the width of the filters and the errors of the observed fluxes are given.}
\label{fig:sed} 
\end{center} 
\end{figure}

The available optical, Two Micron All Sky Survey \citep[2MASS;][]{skrutskie2006}, and Galactic Legacy Infrared Mid-Plane Survey Extraordinaire (GLIMPSE) Infrared Array Camera (IRAC) data \citep{churchwell2009} already allow one to construct a SED from 0.36 to 8\,$\mu$m (see Fig. \ref{fig:sed}). To extend the SED more to the ultraviolet range, we have analysed an observation in the \textit{UVM2} band ($\sim$\,2200\,\AA) of the UV/Optical Telescope \citep[UVOT;][]{poole2008} on board of the \textit{Swift} satellite. We have applied aperture photometry with the UVOT specific tasks in the {\small{HEASOFT}} package (version 6.15) and obtained 17.60$\pm$0.05\,mag (AB).
 
All data were dereddened using the derived $E(B-V)$, $R_V=3.1$, and the extinction curve model by \citet{fitz1999}. The excellent agreement with the model atmosphere confirms the adopted reddening and thus the spectroscopic temperature. To fit the fluxes of the derived stellar atmosphere from Section \ref{sect:specanalysis} to the photometry, one of the coupled parameters (distance and stellar radius) has to be fixed. Owing to the large uncertainties in the stellar parameters, especially in \logg, Stock~16-12 can cover radii between $\sim$\,1.5--4~\textit{R}/R$_{\odot}$ according to evolutionary models by \citet{lagarde2012} [using the zero-age main sequence (ZAMS) as upper limit for gravity]. With these extrema we obtain distances from the SED of 1.0 and 2.8~kpc, respectively. In contrast, the mean cluster distance of 1.9~kpc requires a radius of $\sim$\,2.75~\textit{R}/R$_{\odot}$ for a proper fit.

What can be immediately recognized in Fig. \ref{fig:sed} is the noticeable IR-excess, starting at 5.8~$\mu$m. The IRAC band at 8~$\mu$m deviates already more than 1\,mag from the theoretical flux. This corresponds to more than three times of the total error budget, calculated using the error propagation by \citet{fitz1999} including $\sigma_{R{_V}}$=0.2 and the individual photometric errors. This is an indication for the presence of circumstellar material around a young (PMS) object. Therefore, a connection with the young environment is most likely. On the other side of the covered wavelength range one can also notice emission in the near-UV, which is in general a sign of accretion. However, the position of the UVOT \textit{UVM2} filter coincides with the UV bump at 2175\,\AA\ of extinction curve models \citep[see e.g.][]{fitz1999}. It is therefore strongly  affected by systematic uncertainties in the extinction law. Furthermore, the strength, position, and width of the UV bump depend on the line of sight. Although UV spectra are preferable, also observations in the additional UVOT bands (\textit{UVW1} and \textit{UVW2}) could provide hints about its strength and the real nature of the possible UV excess. The deviation observed at UV wavelengths could be also explained by an underestimation of the effective temperature or an overestimation of the interstellar reddening. Note however that a decreased reddening would lead to a worse fit of the other available photometry and that the excess in the UV is still present when using a higher effective temperature, i.e. 8800\,K. 

We also applied the SED fitting tool by \citet{robit2007}, which was especially built to deal with young stellar objects and allows to fit all parameters simultaneously\footnote{Note that the UVOT filters are currently not included.}. By permitting extreme ranges for the distance (\textit{d}\,$\le$\,5\,kpc) and extinction ($A_{V}$\,$\le$\,5\,mag), the best model fit was achieved with a total $\chi^2$\,=\,21 (14 degrees of freedom), resulting in the parameters $A_{V}$\,=\,1.23 (including 0.1\,mag circumstellar extinction), \textit{d}\,=\,2.4\,kpc, \Teff\,=\,8976\,K, and \textit{R}\,=\,3.3\,R$_{\odot}$ (their model ID 3017872). The extinction value is close to the one adopted above, the distance is in reasonable agreement to that reported for Stock~16, and the temperature only slightly above the spectroscopic range (\Teff=8400$\pm$400\,K). The latter can be explained by the somewhat higher reddening and to some extent also by the use of solar metallicity models in the fitting tool.

\begin{figure}
\begin{center}
\includegraphics[width=85mm,clip]{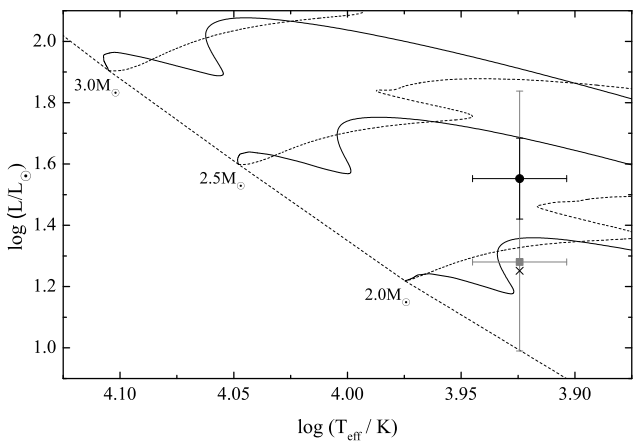}
\caption{The position of Stock~16-12 in the HRD using the $V$-band magnitude, cluster distance and spectroscopic temperature (black circle and error bars), and corrected for binarity with $q=1$ (cross). Furthermore, the purely spectroscopic results were transformed to the temperature/luminosity plane (grey square and error bars). The PMS tracks for solar metallicity with standard prescriptions by \citet{lagarde2012} are indicated as solid lines, whereas ZAMS, MS, and post MS tracks are given as comparison by dashed lines.}
\label{fig:hrd} 
\end{center} 
\end{figure}

In consideration of all available information we therefore conclude that a foreground position of the target star can be excluded and a membership to Stock~16 or the young association is instead very likely. This is also justified by the reasonable agreement of the positions in the Hertzsprung-Russell diagram (HRD) using the cluster distance and the purely spectroscopic results (Fig. \ref{fig:hrd}). We would like to point out that a correction of the object's brightness by 0.75\,mag (binarity with mass ratio $q=1$) results in an exceptional good coincidence with the spectroscopic luminosity value (indicated by the cross in Fig. \ref{fig:hrd}). However, a binary nature is unlikely as discussed in Section \ref{sect:specanalysis}. Nevertheless, the position above the ZAMS supports the PMS nature inferred from the IR-excess, especially if considering the more precise cluster distance. The interpolation within the PMS evolutionary tracks by \citet{lagarde2012} results in 2.3\,$\pm$\,0.2\,M$_\odot$ and an age of 4\,$\pm$\,1\,Myr for the `photometric' position in the HRD, whereas the purely spectroscopic one indicates 1.95\,M$_\odot$ and 6\,Myr. In the last case, the uncertainty especially in age is very large, reaching $\sim$\,15\,Myr towards the ZAMS. The derived ages are well in line with the age of Stock~16 listed in the literature (3--8\,Myr).

%
%
\section{Discussion}
\label{sect:discussion}
In the previous sections we have shown that Stock~16-12 is an intermediate-mass PMS star, which already shows all abundance characteristics of a CP Am star.

The estimated stellar age of 4--6\,Myr is of extreme importance because it allows one to put an observational upper limit on the time-scale of diffusion processes. To our knowledge, the only work which presented stellar models with atomic diffusion for PMS stars is \citet{Vick11}. The earliest appearance of noticeable surface abundance anomalies strongly depends on the mass, whereas low mass loss rates influence the amplitude of peculiarities. Previous calculations, able to reproduce the observed abundances, have already shown that low mass-loss rates ($<10^{-13}\,$M$_{\odot}$\,yr$^{-1}$) for Am stars are likely \citep{michaud1983, Vick11}. 

The evolution of Ca abundance presented by \citet{Vick11} is probably of particular interest, since this element shows a strong variation with age. As can be recognized in their fig. 1, the higher mass models ($\geq$\,1.9\,M$_{\odot}$) predict a short duration ($<$\,1\,Myr) of underabundance, followed by an abrupt change to even significant overabundances. The underabundance dips are located at about 3.5 and 6.5\,Myr for the 2.5 and 1.9 \,M$_{\odot}$ models, respectively. These ages are well in line with our results for the mass, age, and the derived Ca underabundance for Stock~16-12. However, there is a strong temperature dependency (see Table \ref{tab:abundances}) for this element, resulting in even slight overabundances at the adopted upper age limit. Nevertheless, also the evolution of other presented elements (e.g. O, Mn, or Ni) is in agreement with the age and abundance pattern of Stock~16-12. 

Probably the least agreement can be found for Fe, which needs, according to the lower mass models, too long to show the measured overabundance of $\sim$0.4\,dex, but the 2.5\,M$_{\odot}$ model shows a small peak ($\sim$\,0.15\,dex) at the same age as Ca discussed above. The Fe abundance at the lower border of the adopted temperature range would also be in reasonable agreement to the model. Also the higher abundance for this element might be reproduced, though with a much lower mass loss rate than $5 \times 10^{-14}\,$M$_{\odot}$\,yr$^{-1}$ used by \citet{Vick11}. However, also a slightly higher initial metallicity of the host cluster (and thus also of the star) can be responsible.

Chemical peculiarities in Am stars seem to depend also on the stellar rotational velocity \citep{fossati2008}, but it is unsure whether this holds also for PMS stars. Nevertheless, it is worth to obtain a hint of the expected rotational velocity at the ZAMS. Based upon the PMS models by \citet{haemmerle2013}, which include the rotational evolution, the equatorial velocity of the object will be increased by about 30 per cent, though their comparison of the models with the measured rotation rates of Herbig Ae/Be objects shows only a poor agreement. Assuming that the measured \vsini\ represents roughly the true rotational velocity, the target star would reach a velocity of about 90\,\kms\ at the ZAMS, which is close to the upper end of the velocity distribution of Am stars \citep{abt95}.

The possible UV excess (see discussion in Section \ref{sect:clustermember}), which could be interpreted as accretion, has also natural consequences on the elemental distribution in the stellar atmosphere. If further observations indicate that accretion is still in process, the rate has to be smaller than $10^{-14}\,$M$_{\odot}$\,yr$^{-1}$ to have no influence on the chemical separation \citep{Vick11}.

Our non-detection of a magnetic field does not rule completely out the presence of a magnetic field, since the star might have been observed at an unlucky phase. Therefore, one cannot exclude that the star is magnetic and that the current Am abundance pattern of Stock~16-12 might change before the star reaches the ZAMS in order to match that of magnetic CP stars. So far, V380 OriA is the only candidate which shows weak Ap/Bp peculiarities \citep{folsom2012} and a magnetic field. Since the star is about twice as massive as Stock~16-12, its evolution is much faster too. Additional spectropolarimetric observations for Stock~16-12 and the detection of more CP  PMS stars would be essential to test this hypothesis.

\section{Conclusion}
\label{sect:conclusion}
We presented here a detailed parameter determination and abundance analysis for the star Stock~16-12. From the spectroscopic data we determined a temperature of 8400$\pm$400\,K, a surface gravity of \logg=4.1$\pm$0.4\,dex, and a microturbulence velocity of \vmic=3.4$^{+0.7}_{-0.3}$\,\kms. Additional spectroscopic data allowed us to conclude that the object is probably a single star. Furthermore, we have not detected a magnetic field using  spectropolarimetric observations obtained at one epoch.

The star's abundance pattern resembles well that of CP metallic-line (Am) stars. All currently available data suggest that the star is a member of the very young (3--8\,Myr) open cluster Stock~16, or at least to belong to the young Cen OB1 association in which the cluster is embedded. Depending on the adoption of either the purely spectroscopic results or the cluster distance, we determined a stellar mass range of about 1.95--2.3\,M$_\odot$. The membership to a very young region, the noticeable IR-excess, and the stellar age of 4--6\,Myr inferred from evolutionary models allow us to conclude that the object is still in its PMS stage. This star is probably the second PMS CP object known so far, and the very first showing typical Am star elemental abundances.

Young CP stars are extremely important to test and constrain the theory of atomic diffusion. Current available stellar evolutionary models are well in line with our results for the stellar mass, age, and the derived elemental abundances. However, the further evolution of Stock~16-12 is unclear, in particular if the star hosts a structured magnetic field. In this case, the star might develop its current Am abundance pattern into that of MS magnetic CP stars before it reaches the ZAMS. Additional spectropolarimetric measurements and the detection of more CP PMS stars are needed to test this hypothesis.

\section*{Acknowledgements}
Based on observations made with ESO telescopes at the La Silla Paranal Observatory under programme ID 091.C-0498 and 292.C-5044.
This research has made use of data, software, and/or web tools obtained from NASA's High Energy Astrophysics Science Archive Research Center (HEASARC), a service of Goddard Space Flight Center and the Smithsonian Astrophysical Observatory. This research was also made possible through the use of the AAVSO Photometric All-Sky Survey (APASS), funded by the Robert Martin Ayers Sciences Fund. Furthermore, we made use of the SIMBAD data base and VizieR catalogue access tool, CDS, Strasbourg, France, and the WEBDA data base, operated at the Department of Theoretical Physics and Astrophysics of the Masaryk University. MN acknowledges the support by the grant 14-26115P of the Czech Science Foundation. KZ acknowledges support from the Fund for Scientific Research of Flanders (FWO), Belgium, under grant agreement G.0B69.13. EP is financed by the SoMoPro II programme (3SGA5916). The research leading to these results has acquired a financial grant from the People Programme (Marie Curie action) of the Seventh Framework Programme of EU according to the REA Grant Agreement No. 291782. The research is further co-financed by the South-Moravian Region. It was also supported by the grant 7AMB14AT015, and the financial contributions of the Austrian Agency for International Cooperation in Education and Research (BG-03/2013 and CZ-09/2014). This work reflects only the author's views and the European Union is not liable for any use that may be made of the information contained therein.

\label{lastpage}

\end{document}